\documentclass[useAMS,usegraphicx,usenatbib]{mn2e}
\def\revtex@jnl{AAS}%

\newcommand\apj{Astrophysical Journal}%
\newcommand\apjl{Astrophysical Journal, Letters}%
\newcommand\apjs{Astrophysical Journal, Supplement}%
\newcommand\aap{Astronomy and Astrophysics}%
\newcommand\mnras{MNRAS}%
\newcommand\pasj{PASJ}%
\newcommand\araa{Annual Review of Astronomy and Astrophysics}
\newcommand\nodata{ ~$\cdots$~ }
\usepackage{amssymb,amsmath}

\newcommand{\xmm}{\it XMM-Newton\rm}
\newcommand{\chandra}{\it Chandra\rm}
\newcommand{\rosat}{\it ROSAT\rm}
\newcommand{\suzaku}{\it Suzaku\rm}

\def\a1835{\emph{Abell~1835}}

\def\rtwo{$r_{200}$}
\def\rfive{$r_{500}$}

\title[Chandra X-ray observations of  Abell~1835 to the virial radius]{Chandra X-ray observations of  Abell~1835 to the virial radius}
\author[M. Bonamente, D. Landry, B. Maughan, P. Giles, M. Joy and J. Nevalainen]{M. Bonamente$^{1,2}$, D. Landry$^{1}$,
 B. Maughan$^{3}$, P. Giles$^{3}$, M. Joy$^{2}$ and J. Nevalainen$^{4}$
\\
$^{1}$Physics Department, University of Alabama in Huntsville,
Huntsville, AL, U.S.A.\\
$^{2}$NASA National Space and Technology Center, Huntsville, AL, U.S.A.\\
$^{3}$HH Wills Physics Laboratory, Tyndall Avenue, Bristol, BS8 1TL, UK\\
$^{4}$University of Helsinki, Finland}

\begin{document}

\date{Accepted . Received ; in original form }

\pagerange{\pageref{firstpage}--\pageref{lastpage}} \pubyear{2012}

\maketitle

\label{firstpage}






\begin{abstract}
We report the first \chandra\ detection of emission out to the
virial radius in the cluster Abell~1835 at $z=0.253$. 
Our analysis of the soft X-ray surface brightness shows that
emission is present out to a radial distance of 10~arcmin or
2.4~Mpc, and the temperature profile has a factor of ten drop
from the peak temperature of 10~keV to the value at the virial
radius.
We model the \chandra\ data from the core to the virial radius and
show that the steep temperature profile is not compatible with
hydrostatic equilibrium of the hot gas, and that the gas
is convectively unstable at the outskirts. 
A possible interpretation of the \chandra\ data is the presence of 
a second phase of \emph{warm-hot}
gas near the cluster's virial radius that is not in hydrostatic equilibrium
with the cluster's potential. The observations are also consistent
with an alternative scenario in which the gas is significantly clumped
at large radii.
\end{abstract}





\section{Introduction}
The large-scale halo of hot gas  provides
a unique way to measure the baryonic and gravitational
mass of galaxy clusters.
The baryonic mass can be measured directly from
the observation of the hot X-ray emitting intra-cluster medium (ICM), and 
of the associated stellar component \citep[e.g.][]{giodini2009,gonzales2007},
while measurements of the gravitational mass require
the assumption of hydrostatic equilibrium between the gas and
dark matter. 
Cluster cores are subject to a variety of non-gravitational
heating and cooling processes that may result in deviations
from hydrostatic equilibrium, and in inner regions beyond the
core the ICM is expected to be in hydrostatic equilibrium
with the dark matter potential. At the
outskirts, the low-density ICM and the proximity to the
sources of accretion results in the onset
of new physical processes
such as  departure from hydrostatic equilibrium \citep[e.g.,][]{lau2009},
clumping of the gas \citep{simionescu2011},
different temperature between electrons and ions \citep[e.g.,][]{akamatsu2011},
and flattening of the entropy profile \citep{sato2012}, leading
to possible sources of systematic uncertainties in the measurement of masses.

The detection of hot gas at large radii is limited
primarily by its intrinsic low surface brightness,  uncertainties
associated with the subtraction of background (and foreground) emission,
and the ability to remove contamination from
compact sources unrelated to the cluster.
Thanks to its low detector background, \suzaku\ reported the
measurement of ICM temperatures  to \rtwo\ and beyond
for a few nearby clusters
\citep[e.g.][]{akamatsu2011,walker2012a,walker2012b,simionescu2011,burns2010,kawaharada2010,
bautz2009,george2009}; to date \a1835\
has not  been the target of a \suzaku\ observation.

In this paper we report the \chandra\ detection of X-ray emission
  in \a1835\ beyond \rtwo, using three observations 
for a total of 193~ksec exposure time, extending the analysis
of these \chandra\ data performed by \cite{sanders2010}.
The radius $r_{\Delta}$ is defined as the radius within which
the average mass density is $\Delta$ times the critical density of
the universe at the cluster's redshift for our choice of 
cosmological parameters. The virial radius of a cluster
is defined as the equilibrium radius of the collapsed
halo, approximately equivalent to one half of its turnaround radius 
\cite[e.g.][]{lacey1993, eke1998}.
For an $\Omega_{\Lambda}$-dominated universe,
the virial radius is approximately $r_{100}$ \citep[e.g.][]{eke1998}.
\a1835\ is the most luminous cluster in the 
\cite{dahle2006} sample of clusters at $z=0.15-0.3$ selected
from the \emph{Bright Cluster Survey}.
The combination of high luminosity and availability of
 deep \chandra\ observations with local background make \a1835\
and ideal candidate to study its emission to the virial radius. 
\a1835\ has a redshift of $z=0.253$,
which for $H_{0}=70.2$~km~s$^{-1}$~Mpc$^{-1}$, $\Omega_{\Lambda}=0.73$,
$\Omega_M=0.27$ cosmology \citep{komatsu2011} corresponds
to an angular-size distance of $D_A=816.3$~Mpc, and a scale of 
237.48 kpc per arcmin.

\section{Chandra and ROSAT observations of Abell~1835 and the detection
of cluster emission beyond \rtwo}
\label{sec:Sx}
\subsection{Chandra observations} 
\begin{figure}
\centering
\includegraphics[width=3.1in,angle=-90]{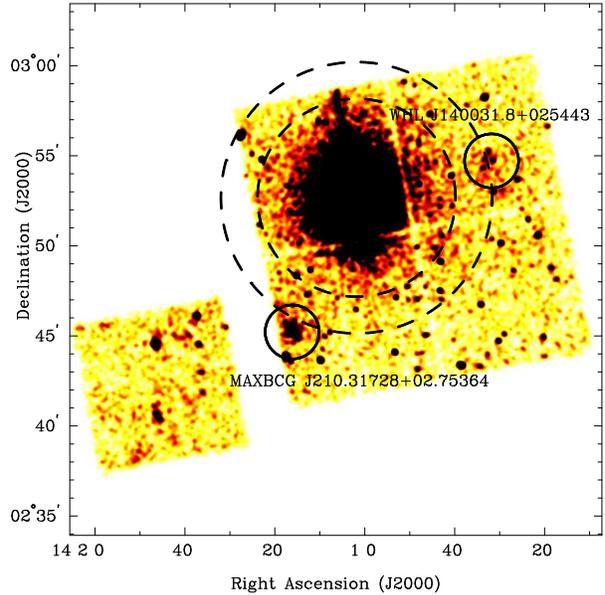}
\caption{Image of \a1835\ from observation 6880, in the 0.7-2 keV band.
The data were smoothed with a Gaussian kernel of $\sim$6 arcsec standard error.
The dashed circles correspond to radial distances of approximately \rfive\ and \rtwo,
and the full black circles mark the position of the two low-mass clusters associated with
\a1835.}
\label{fig:a1835}
\end{figure}

\chandra\ observed \a1835\ three times between December 2005 and August 2006
(observations ID 6880, 6881 and 7370), with a combined clean exposure time
of 193~ks. The three observations had similar aimpoint towards the
center of the cluster (R.A. 14h01m02s, Dec. +02d51.5m J2000) and different
roll angles. All observations were taken with the ACIS-I detector configuration,
which consists of four ACIS front-illuminated chips in a two-by-two square,
plus a fifth identical chip that may be used to measure 
the \emph{in situ} soft X-ray background.
Figure~\ref{fig:a1835} is an image from the longest observation (ID 6880, 118ks)
in the soft X-ray band (0.7-2 keV). In addition to a large number of 
compact X-ray sources that were excluded from further analysis, the data
show a clear detection of diffuse X-ray emission associated with two
additional low-mass clusters identified from the \emph{Sloan Digital Sky Survey}, 
MAXBCG J210.31728+02.75364 and WHL J140031.8+025443.
The cluster MAXBCG J210.31728+02.75364 is the only cluster in the vicinity of \a1835\
reported in the MAXBCG catalog of \cite{koester2007}, and it has
a measured  photo-$z$
of 0.238, while the catalog of \cite{wen2009} reports a photo-$z$
of 0.269 for the same source; given the uncertainties associated
with photometric redshifts, it is likely that the cluster is
in physical association with \a1835\ ($z=0.253$).
The \cite{wen2009} catalog also reports another optically-identified cluster 
in the area, WHL J140031.8+025443, with a spectroscopic redshift of $z=0.2505$.
The association of these two groups with \a1835\ is confirmed by redshift
data provided by C. Haines (personal communication), who measures
a redshift of $z=0.250$ for WHL J140031.8+025443, and $z=0.245$ for MAXBCG J210.31728+02.75364.

Since the goal of this paper is to study the
diffuse emission associated with \a1835, we excise a region of radius 90~arcsec
around the position of the two clusters (black circles in 
Figure~\ref{fig:a1835}), and study their emission separately from that of \a1835\
(see Section~\ref{sec:low-mass-clusters}).

\subsection{Chandra data analysis}
The reduction of the \chandra\ observations follow the procedure
described in \cite{bonamente2006} and \cite{bonamente2011}, which consists of filtering the
observations for possible periods of flaring background,
and applying the latest calibration; no significant flares
were present in these observations. The reduction was
performed in CIAO~4.2, using CALDB 4.3; in Sec.~\ref{sec:robustness} we discuss the impact of calibration
changes on our results. One of the calibration issues
that can affect the measurement of cluster emission is the uncertainty in
the contamination of the optical blocking filter, which causes
a reduction in the low energy quantum efficiency of the \chandra\ detectors.
The spatial and time dependence of this contaminant affects primarily
the effective area at $\leq$0.7 keV~\footnote{See \cite{marshall2004} and 
\chandra\  calibration memos at cxc.harvard.edu.}, with an estimated residual
error of $\leq$ 3\% at higher energy. We therefore limit our spatial and spectral 
analysis to the $\geq$0.7~keV band.
The superior angular resolution of the \chandra\ mirrors \citep{weisskopf2000}
results in a point-spread function with a 0.5~arcsec FWHM, and therefore 
there is negligible contribution from the bright cluster core to 
the emission in the outer annuli, and from secondary scatter (stray light) by sources outside
the field of view.

The subtraction of particle and sky background is one of the
most crucial aspects of the analysis of low surface brightness cluster regions.
We use \chandra\ blank-sky background observations,
rescaled according to the high-
energy flux of the cluster, to ensure a correct subtraction
of the particle background that is dominant at $E\geq9.5$~keV, 
where the Chandra detectors have no effective area.
The temporal and spatial variability of the soft X-ray background at $E<2$~keV
also requires that a peripheral region free of cluster emission
is used to measure any local enhancement (or deficit) of soft X-ray
emission relative to that of the blank-sky fields, and account for this
difference in the analysis. This method 
is accurate for the determination of the temperature profile, but may result
in small errors in the measurement of the surface brightness profile.
In fact, the blank-sky background is a combination of a particle component 
that is not vignetted, and a sky component that is vignetted. To determine the
surface brightness of the cluster and of the local soft X-ray background,
a more accurate procedure consists of subtracting the non-vignetted
particle component as measured from \chandra\ observations in which
the ACIS detector was stowed \cite[e.g.,][]{hickox2007}, after rescaling the
stowed background to match the $E\geq9.5$~keV cluster count rate, as in 
the case of the blank-sky background.

Point sources are identified and removed using a wavelet detection method
that correlates the cluster observation with wavelet functions of 
different scale sizes (\emph{wavdetect} in CIAO). Subtraction
of point soures from the blank-sky observations were 
performed by eye, with results that closely match those
of the wavelet method.

\begin{figure}
\centering
\includegraphics[width=2.5in, angle=-90]{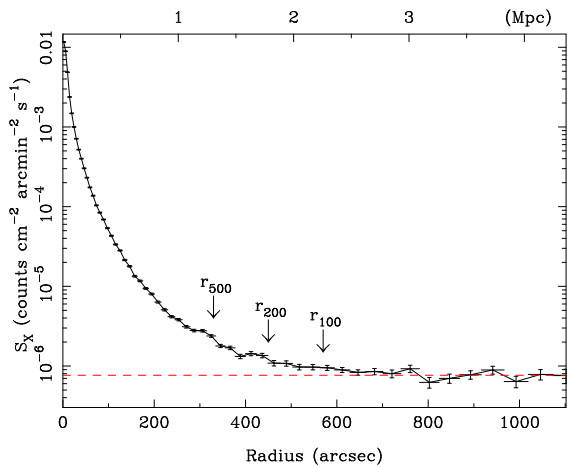}
\caption{Exposure corrected surface brightness profile of \a1835\ in the soft X-ray band (0.7-2 keV),
obtained by subtraction of the particle background from the ACIS stowed
observations. The radii \rfive, \rtwo\ and the virial radius ($\sim r_{100}$) are 
estimated from the data in Section~\ref{sec:r500} (see Table~\ref{tab:vikh-masses}).
The dashed red line is the average background level in the region $\geq$ 700 arcsec.}
\label{fig:Sx}
\end{figure}

\subsection{Measurement of the surface brightness profile with Chandra}
\begin{figure}
\centering
\includegraphics[width=2.5in, angle=-90]{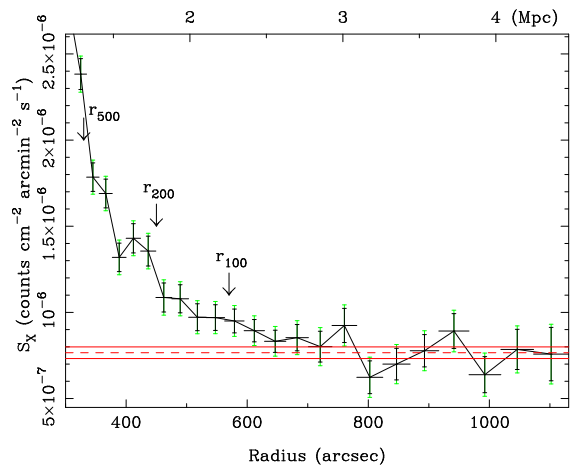}
\caption{Close-up view of Figure~\ref{fig:Sx}, in which the red lines represent the
1-$\sigma$ confidence in the background level as determined from the $\geq$700~arcsec region,
and the green error bars combine the statistical and systematic errors in the determination
of the surface brightness.
}
\label{fig:Sx-closeup}
\end{figure}

The surface brightness profile obtained using this background subtraction
is shown in Figure~\ref{fig:Sx}, in which the red line represents
the average value of the background at radii $\geq$~700 arcsec, where
the surface brightness profile is consistent with a constant level.
To determine the outer radius at which \chandra\ has a significant detection
of the cluster, we also include sources of systematic errors in our analysis.
One source of uncertainty is the error in the measurement of the background level,
shown in Figure~\ref{fig:Sx-closeup} as the solid red lines.
The error is given by the standard deviation of the weighted mean of the datapoints
at radii greater than 700~arcsec,
 to illustrate that
each bin in the surface brightness profile beyond this radius
is consistent with a constant level of the background. 

Another source of uncertainty is the amount by which the stowed background is to
be rescaled to match the cluster count rate at high energy. The stowed background
dataset applicable to the dates of observation of \a1835\ has an exposure
time of 367~ksec, and the relative error in the rescaling of the background to match
the cluster count rate at high energy is 0.7\%, as determined by the Poisson
error in the photon counts at high energy. 
Moreover, \cite{hickox2006}
has shown that the spectral distribution of the particle background is remarkably stable,
even in the presence of changes in the overall flux, and that 
the ratio of soft-to-hard (2-7 keV to 9.5-12 keV) count rates remains constant to within $\leq$2~\%. We therefore
apply a systematic error of 2~\% in the stowed background flux, to account for this
possible source of uncertainty, in addition to the 0.7\% error
due to the uncertainty in the rescaling of the background. 

In Figure~\ref{fig:Sx-closeup} the green error
bars represent the cumulative effect of the statistical error due to the counting
statistic,  and the
sources of errors associated with the use of the stowed background; the systematic
errors were added linearly to the statistical error as a conservative measure.
This error analysis shows that the emission from \a1835\ remains significantly 
above the background beyond \rtwo\ and
until approximately a radius of 600 arcsec, or approximately 2.4~Mpc. 
The significance of the detection in the region 450-600" (the
five datapoints in Figure~\ref{fig:Sx-closeup} after the $r_{200}$ marker)
is calculated as 5.5$\sigma$, and is obtained by using the larger systematic
error bars for the surface brightness profile (in green in Figure~\ref{fig:Sx-closeup}),
added in quadrature to the error in the determination of the background level
from the $\geq 700$" region (red lines in Figure~\ref{fig:Sx-closeup}).

To further test the effect of the background subtraction, we repeat our
backround subtraction process using the $\geq 600$" region 
(instead of the $\geq 700$" region) . The background level
increases by less than 1$\sigma$ of the value previously determined (e.g., the
two levels are statistically indistinguishable), and the significance of detection in
the region 450-600" is 4.7$\sigma$. Therefore we conclude that it is unlikely that
the excess of emission beyond \rtwo\ and out to the virial radius is due to errors in the background
subtraction process.
A similar result can be obtained including the 2-7 keV band,
but the signal-to-noise is reduced because at large radii this band is 
dominated by the background due to the softening of the cluster emission. 
We estimate \rtwo\ and the virial radius ($\sim r_{100}$) from
the \chandra\ data in Section~\ref{sec:r500}.

\subsection{Measurement of the
surface brightness profile with the ROSAT Position Sensitive Proportional Counter}

\rosat\ observed \a1835\ on July 3--4 2003 for 6~ks with the Position
Sensitive Proportional Counter (PSPC), observation ID was 800569.
The PSPC has a 99.9\% rejection of particle background in the 0.2-2~keV band
\citep{plucinsky1993} and an average angular resolution of $\sim$30~arcsec that makes it 
very suitable for observations of low surface brightness objects such as
the outskirts of galaxy clusters \citep[e.g.][]{bonamente2001,bonamente2002,bonamente2003}.
We reduce the event file following the procedure described in  \cite{snowden1994}
and \cite{bonamente2002}, which consists of corrections for detector gain fluctuations, and
removal of periods with a \emph{master veto} rate of $\leq$170 counts~s$^{-1}$ in order
to discard periods of high background. These filters result in a clean exposure time of
5.9~ks.

\begin{figure}
\includegraphics[width=2.4in,angle=-90]{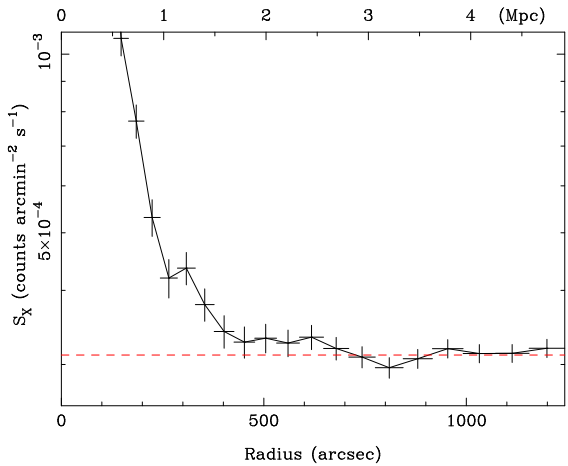}
\caption{Surface brightness profile in 0.2-2 keV band from a 6~ks
observation with ROSAT PSPC. The background level is determined from 
the data at radii $\geq$~700", as in the \chandra\ data.}
\label{fig:rosat}
\centering
\end{figure}

Since the PSPC background is given only by the photon background, we generate an
image in the 0.2-2 keV band and use the exposure map to correct for the position--dependent 
variations in the detector response and mirror vignetting.
We masked out the two low-mass cluster regions as we did for the \chandra\ data and 
all visible point sources, and obtained and exposure-corrected surface brightness profile
out to a radial distance of $\sim$20 arcmin, which corresponds to the location of the
inner support structure of the PSPC detector. The \rosat\ surface brightness profile therefore
covers the entire azimuthal range.
In Figure~\ref{fig:rosat} we  show the radial profile of the surface brightness in the 0.2-2 keV
band, showing a $\sim$2~$\sigma$ excess of emission in the 400-600" region using the background
level calculated from the region $\geq$700", as done for the \chandra\ data.
The \rosat\ data therefore provide additional evidence of emission beyond \rfive\ and out to the
virial radius, although the short \rosat\ exposure does not have sufficient number of counts
to provide a detection with the same significance as in the \chandra\ data.

\section{Analysis of the Chandra spectra}
\label{sec:kT}
\subsection{Measurement of the temperature profile of Abell~1835}
\label{sec:spectral-fits}
\begin{figure}
\centering
\includegraphics[width=2.3in,angle=-90]{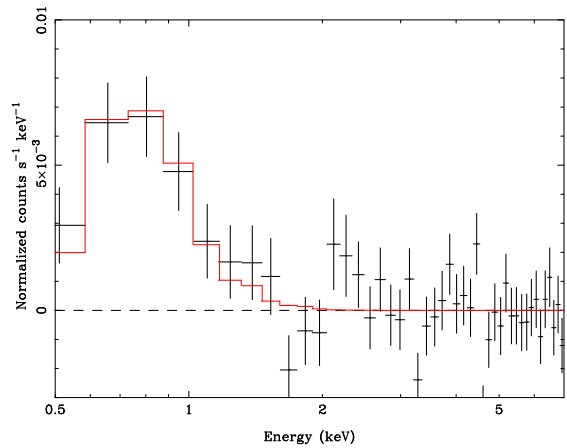}
\caption{Spectrum of the local enhancement of the soft X-ray
background from observation 6880. The other two exposures have
similar levels of soft X-ray fluxes above the blank-sky emission,
which is modeled as an unabsorbed $\sim0.25$~keV thermal plasma at $z=0$.
The best-fit model has a $\chi^2_{min}=73.9$ for 78 degrees of freedom,
for a null hypothesis probability of 61\%.}
\label{fig:soft-back}
\end{figure}

We measure the temperature profile of \a1835\ following
the background subtraction method described in Sec.~\ref{sec:Sx},
which makes use of the blank-sky background dataset and a
measurement of
the local enhancement of the soft X-ray background, as
is commonly done for \chandra\ data \citep[e.g.][]{vikhlinin2006, maughan2008, bulbul2010}.
In Figure~\ref{fig:soft-back} we show the spectral distribution
of the local soft X-ray background enhancement, as determined from
a region beyond the virial radius ($\geq$700~arcsec);
this emission was modelled with an APEC emission
model of $kT\sim 0.25$~KeV and of Solar abundance, consistent with
Galactic emission, and then subtracted from all
spectra. The spectra were fit in the 0.7-7 keV band using the 
minimum $\chi^2$ statistic, after binning to ensure that there are
at least 25 counts per bin. We use XSPEC version 12.6.0s
for the spectral analysys.

\begin{figure*}
\centering
\includegraphics[width=2.5in,angle=-90]{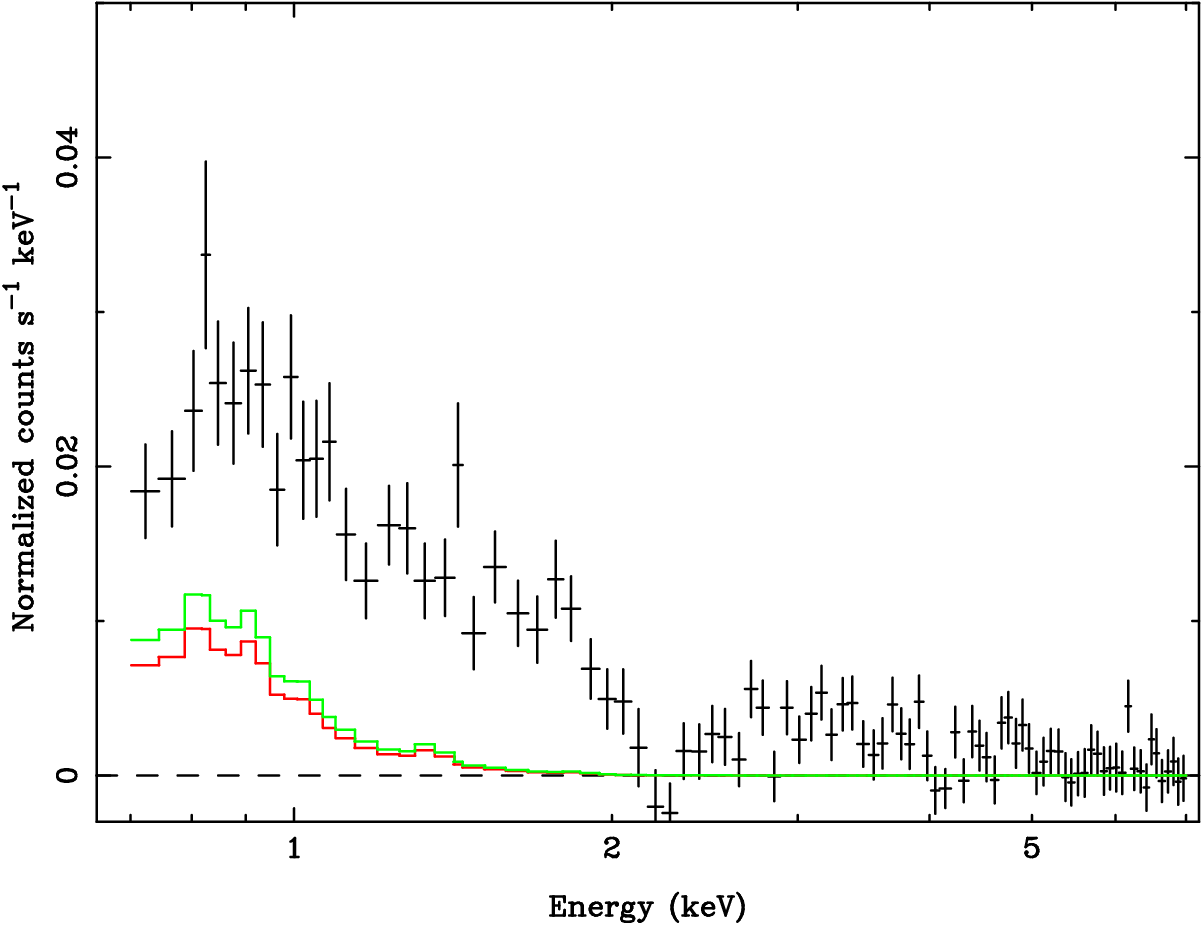}
\includegraphics[width=2.5in,angle=-90]{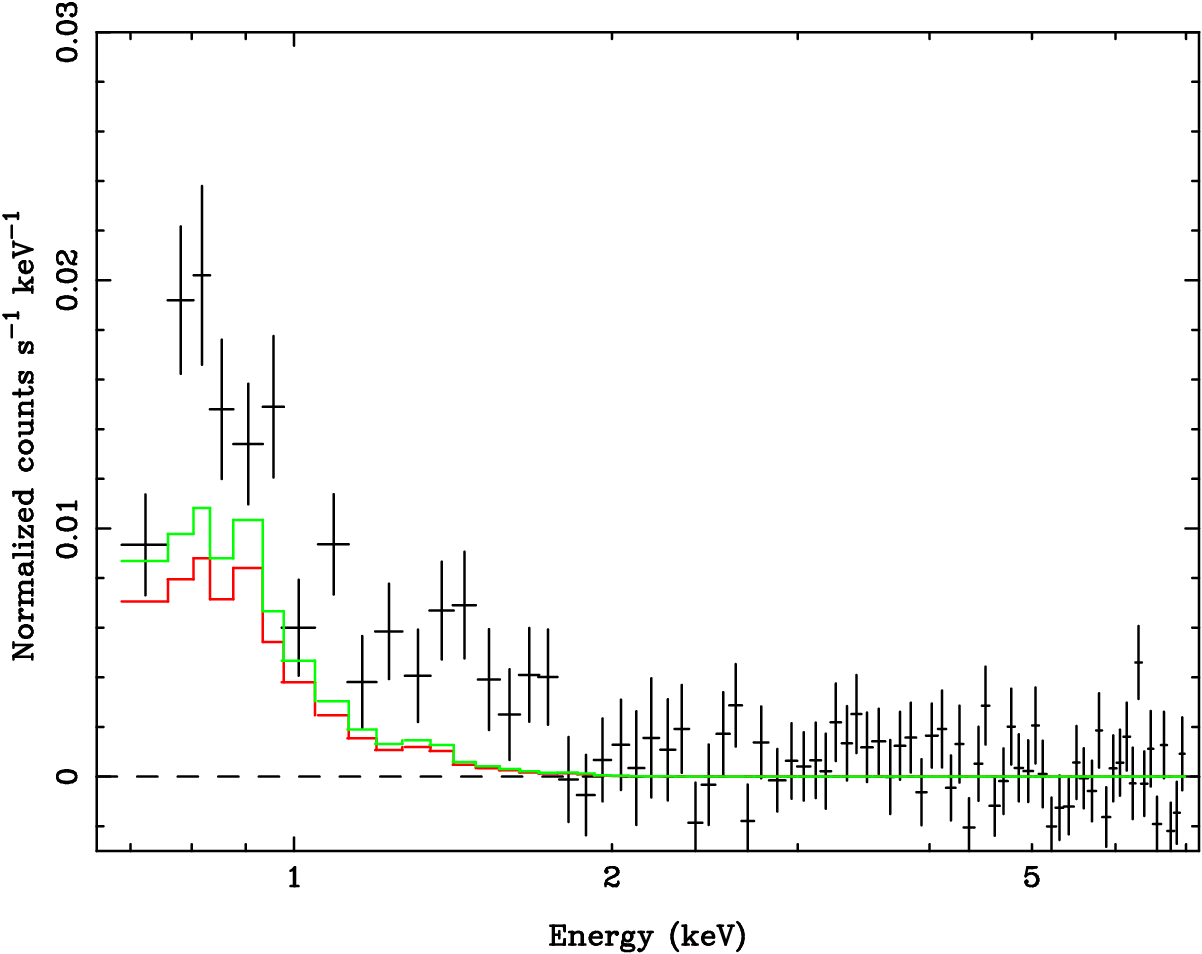}
\caption{Blank-sky background subtracted spectra of
regions 330-450" and 450-600" from observation 6880.
The solid lines are the best-fit model of the local
soft X-ray enhancement of Figure~\ref{fig:soft-back} (red),
and its 90\% upper limit (green).}
\label{fig:spectra}
\end{figure*}

In Figure~\ref{fig:spectra} we show the spectra of
the outermost two regions, to show the impact of the 
soft X-ray residuals in the background subtraction.
The importance of background systematics in the detection
of emission and measurement of cluster temperatures for
regions of low surface brightness was
recently addressed by \cite{leccardi2008} using \xmm\ data.
For our \chandra\ observations, the two main sources of uncertainty when determining the temperature
of the outer regions are the subtraction of the blank-sky background, and
the subtraction of the locally-determined soft X-ray background.
Table~\ref{tab:background} reports the statistics of the background relevant
to the outer regions of the cluster, with  both regions $\sim$~10-20\% above
the blank-sky background,  determined with a precision of 1-2\%.
The additional soft X-ray background accounts for a significant portion
of the remaining signal, as shown in Figure~\ref{fig:spectra}; the 90\% 
upper limit to the measurement  of this background is shown as the
green lines, and emission from the cluster is still detected with
high statistical significance. Both sources of error are 
included in the temperature measurements at large radii.

\begin{table*}
\centering
\caption{Background levels in outer regions}
\label{tab:background}
\begin{tabular}{lccc}
\hline
 		& \multicolumn{3}{c}{Observation ID}\\
	 	& 6880 	& 6881	&7370\\ 
\hline
Exposure time (ks) & 114.1 & 36.0 & 39.5 \\
Correction to Blank-sky Subtraction$^{a}$ & -0.04$\pm$0.01 & -0.125$\pm$0.015 & -0.04$\pm$0.015 \\
\hline
\multicolumn{4}{c}{Region 330-450"}\\
Total Counts	& 18,124 & 4,938 & 5,686 \\
Count rate (c s$^{-1}$) & 0.158$\pm$0.001 & 0.137$\pm$0.002 & 0.144$\pm$0.002\\
Net count rate$^{b}$ ($10^{-2}$ c s$^{-1}$) 	& $2.75\pm0.10$ & $2.54\pm0.20$ & $2.36\pm0.20$ \\
Percent above back. & 17.4$\pm$0.6 & 18.2$\pm$1.4 & 16.4$\pm$1.4 \\
SXB count rate ($10^{-3}$ c s$^{-1}$) 	&  3.34$\pm$0.77 & 7.20$\pm$0.94 & 7.10$\pm$0.71 \\
\hline
\multicolumn{4}{c}{Region 450-600"} \\
Total Counts    & 15,811 & 4,901 & 5,483 \\
Count rate (c s$^{-1}$) & 0.139$\pm$0.001 & 0.136$\pm$0.002 & 0.139$\pm$0.002 \\
Net count rate$^{b}$ ($10^{-2}$ c s$^{-1}$)     &1.02$\pm$0.10 & 1.58$\pm$0.20 & 1.23$\pm$0.20 \\
Percent above back. & 7.3$\pm$0.7 & 11.6$\pm$1.5 & 8.8$\pm$1.4 \\
SXB count rate ($10^{-3}$ c s$^{-1}$)   & 3.06$\pm$0.70 & 7.50$\pm$0.98 & 7.32$\pm$0.73\\
\hline
\end{tabular}

\flushleft
$a$: This is the fractional correction of the blank-sky data, to match the high-energy flux
in the cluster observation.   \\
$b$ This is the background-subtracted count rate, including cluster and soft X-ray background 
(SXB) signal.
\end{table*}

We use the APEC code \citep[][code version 1.3.1]{smith2001} to model the \chandra\ spectra, with a
fixed Galactic HI column density of $N_H=2.04\times10^{20}$~cm$^{-2}$ \citep{kalberla2005}.
The region at radii $\leq 330$" have a variable metal abundance, while the outer
region have a fixed abundance of $A=0.3$. In addition to the statistical errors
obtained from the XSPEC fits, we add a systematic error of 10\% in the
temperature measured in the core and a 5\% error to the other region,
to account for possible systematic uncertainties due to the \chandra\
calibration \citep[see, e.g.,][]{bulbul2010}. 
One possible source of systematic uncertainty in our results is indicated 
by the systematic difference between the \chandra/ACIS and\xmm/EPIC 
temperature measurements of galaxy clusters \citep{nevalainen2010}
This amounts to a $\pm$10\% bias in the calibration of the effective area at 0.5 keV, 
which decreases roughly linearly towards 0\% bias at 2 keV. 
Assuming that \xmm/pn has a more accurately calibrated effective area, 
we reduced the \chandra\ effective area by multiplying it with a linear function 
as indicated by the \chandra/\xmm\ comparison. As a result, the temperature at the 
outermost radial bin decreases by $\sim$ 5\%. Thus, the cross-calibration 
uncertainties between \chandra\ and \xmm\ do not explain the low temperature we measure in the outermost radial bin.
Uncertainties in the Galactic column density of HI do not impact significantly our results.
Changing the value of  $N_H$ by $\pm$10\%, consistent with the variations between the \cite{kalberla2005}
and the \cite{dickey1990} measurements, results in a change of best-fit temperature
in each bin by less than 2\%.

Given the emphasis of this paper on the detection of
emission at large radii, we investigate
the sources of uncertainty caused by the background subtraction
in the outer region at $\geq$330". We report the results of this 
error analysis in Table~\ref{tab:kT-err}, where \emph{cornorm} refers
to the normalization of the blank-sky background, and \emph{soft residuals}
refers to the normalization of the soft X-ray residual model, as 
reported in Table~\ref{tab:background}. In the analysis that follows,
we add the systematic errors caused by these sources linearly to the
statistical error.
Our data do not constrain well the metal abundance of the plasma
in the outer regions.
Using an abundance of $A=0.5$ instead of the nominal $A=0.3$
leads to negligible changes in the best-fit temperature for both of
the outer annuli.
In the extreme case of an $A=0.0$ metal abundance, 
both regions have an acceptable fit with the best-fit temperatures
change respectively by $+6$\% for the 330-450" region ($\Delta \chi^2=+1.3$), and
by $-22$\% for the 450-600" region ($\Delta \chi^2=+9.2$,
best fit decreases from 1.26 to 0.98 keV). 
We therefore find that, in the case of exceptionally low metalllicity, the
temperature profile we measure from these \chandra\ data would be even significantly steeper
than indicated by the result in Table~\ref{tab:kT-err}.
Given that these data do not provide direct indication that the plasma in the outer regions 
may have null metal content, we do not fold in this source of systematic error in the
analysis that follows.

\begin{table}
\caption{Temperature measurement and error analysis from
the \chandra\ data.}
\centering
\label{tab:kT-err}
\begin{tabular}{lcc}
\hline
Region & \multicolumn{2}{c}{Projected Temperature (keV)}\\
\hline
       & Measurement$^{a}$  & Calibration error$^{b}$  \\
0-10"  & 4.78$\pm$0.06 & $\pm$0.48 \\
10-20" & 7.09$\pm$0.14 & $\pm$0.71 \\
20-30" & 8.72$\pm$0.27 & $\pm$0.87 \\
30-60" & 9.47$\pm$0.21 & $\pm$0.47 \\ 
60-90" & 10.57$\pm$0.33 & $\pm$0.53 \\
90-120" & 9.97$\pm$0.44 & $\pm$0.50 \\
120-180"& 9.68$\pm$0.49 & $\pm$0.48 \\
180-240" & 7.85$\pm$0.65 &$\pm$0.39 \\
240-330" & 6.02$\pm$0.65 &$\pm$0.30 \\
330-450" & 3.75$\pm$0.72 & $\pm$0.19\\
450-600" & 1.26$\pm$0.16 & $\pm$0.06 \\
\hline
	\multicolumn{3}{c}{Measurement of Temperature Using} \\
	 \multicolumn{3}{c}{Background Systematic Errors (keV)}\\
         & $+1\sigma$ cornorm$^{c}$ & $-1\sigma$ cornorm \\
330-450" & 3.02$\pm$0.54 & 4.67$\pm$1.00 \\
450-600" & 1.09$\pm$0.10 & 1.31$\pm$0.18 \\
	 & $+1\sigma$ soft res.$^{d}$ & $-1\sigma$ soft res.\\
450-600" & 4.53$\pm$1.03 & 3.05$\pm$0.54\\
450-600" & 1.37$\pm$0.25 & 1.16$\pm$0.12\\
 \multicolumn{3}{c}{Summary of Background Systematic Errors$^{e}$} \\
330-450" & \multicolumn{2}{c}{$\pm 0.83\pm 0.74$ keV} \\ 
450-600" & \multicolumn{2}{c}{$\pm 0.11\pm 0.10$ keV} \\
\hline
\end{tabular}

\flushleft
$a$: Uncertainty is 1$\sigma$ statistical error from counting statistics only.\\
$b$: Includes \xmm/\chandra\ cross-calibration uncertainty of the effective area \citep{nevalainen2010}.\\
$c$: This is temperature obtained by varying by $\pm 1\sigma$ the fractional
correction of the blank-sky data, to match the high-energy flux in the cluster observations.\\
$d$: This is the temperature obtained by varying by $\pm 1\sigma$ the normalization
of the best-fit model to the soft X-ray background residuals.\\
$e$: Obtained from the average deviation of the $\pm 1\sigma $ `cornorm' and 'soft. res'
measurements from the measurement with nominal values of these parameters.
\end{table}

\cite{sanders2010} measured temperature profiles for \a1835\ out
to approximately \rfive\ with \chandra\ and \xmm. Using the same \chandra\ 
observations we analyze in this paper, their temperature profile
has a similar drop from the peak value to their outermost
annulus ($322\pm42$"), where they measure a temperature of kT=$4.67\pm^{0.82}_{0.52}$~keV
that is consistent with our measurements. Likewise, from the \xmm\ data their outermost
radial bin ($300\pm10$") has a temperature of kT=$5.2\pm^{1.2}_{0.7}$~keV, also
in agreement with our results. 
The only measurement of the \a1835\ temperature to the virial radius
available in the literature is that of \cite{snowden2008}, who
does report a temperature profile out to a distance of 12~arcmin
from a long \xmm\ observation (and out to 7' from a shorter observation).
In particular,
they report a temperature of $kT=3.14\pm0.93$ for the region
420-540", which straddles our measurements at 330-450" ($3.75\pm0.72$~keV)
and at 450-600" ($1.26\pm0.16$, statistical errors only). The same paper
also reports a measurement of $kT=3.33\pm1.75$~keV for the region 540-720",
i.e., beyond our outer annulus. Their temperature is somewhat higher that
ours, although the large error bars cannot exclude that the \chandra\
and \xmm\ measurements
are consistent. Therefore our results confirm and extend the
earlier \xmm\ analysis of \cite{snowden2008}.

\subsection{Measurement of the average temperature of 
MAXBCG J210.31728+02.75364
and WHL J140031.8+025443}
\label{sec:low-mass-clusters}
We also measure the temperature of the two SDSS clusters detected
in our \chandra\ images, 
MAXBCG J210.31728+02.75364 
and WHL J140031.8+025443. The two clusters  are located between a distance of $\sim$380-650"
from the cluster center, and therefore we start by extracting a spectrum for this annulus
excluding two regions of 1.5' radius centered at the two clusters.
This radius was determined by visual inspection, after smoothing of the \chandra\ image with a
Gaussian kernel of $\sigma=6$ arcsec. 
For this annulus, we measure a temperature of $kT=1.85\pm0.36$~keV for a fixed
abundance of $A=0.3$ Solar. We then use this spectrum as the local background for the
two cluster regions, and measure a temperature $kT=2.73\pm^{0.93}_{0.54}$~keV i
for MAXBCG J210.31728+02.75364
(357 source photons, 19\% above the average emission of the annulus),
and $kT=2.09\pm^{4.6}_{0.55}$~keV for WHL J140031.8+025443 (538 photons, 27\% above background). 
For both clusters, we assumed the same Galactic $HI$ column density as for \a1835, and
a fixed metal abundance of $A=0.3$ Solar. For both clusters we also extract spectra in
regions larger than 1.5', and determine that no additional source photons are present
from these two clusters beyond this radius.

\subsection{Tests of robustness of the temperature measurement at large radii}
\label{sec:robustness}
To further test the measurement of temperatures especially at large radii,
where
the background subtraction is especially important,
we also measure the temperature profile using the same stowed
background data that was used for the surface brightness measurement
of Figures~\ref{fig:Sx} and \ref{fig:Sx-closeup}. As in the case
of the blank-sky background, we first rescale the stowed data to match
the high-energy count rate of the cluster observation, and
use a region at large radii ($\geq$ 700 arcsec) to measure
the local X-ray background. We model the background using
an APEC plus a power-law model, the latter component necessary
to model the harder emission due to unresolved AGNs that is typically removed when
the blank-sky background is used instead, and apply this
model to all cluster regions. We find that the temperature profile
is consistent within the 1~$\sigma$ statistical errors 
of the values provided in Table~\ref{tab:kT-err} for each region,
and therefore conclude that the temperature drop at large radii, and
especially in the outermost region, is not sensitive
to the background subtraction method.

The temperature measurement is also dependent on an accurate subtraction
of background (and foreground) sources of emission. 
Point sources in the field of view are detetected using the CIAO
tool  \emph{wavdetect},  which correlates the image with wavelets
at small angular scales (2 and 4 pixels, one pixel is 1.96"), searches the results for 3-$\sigma$ correlations, and
returns a list of elliptical regions to be excluded from the analysis. 
We study in particular the
effect of background sources on the measurement of the temperature in the outermost
annulus (450-600"). In this region,  \emph{wavdetect} finds
24 point sources, plus portions of the two low-mass galaxy clusters
described in Section~\ref{sec:low-mass-clusters}.
We extract a spectrum for this region from the longest observation (ID 6880),
and now include in the spectrum all
point sources excluded in the previous analysis.
We find a count rate of $3.20\pm0.13\times 10^{-2}$ counts~s$^{-1}$, compared
to the point source-subtracted rate of $1.02\pm0.10\times 10^{-2}$ counts~s$^{-1}$,
corresponding to an increase in background-subtracted flux by a factor of three.
We then fit the spectrum with the same APEC model as described in Section~\ref{sec:spectral-fits},
and find a best-fit temperature of $kT=1.96\pm0.17$~keV for a best-fit goodness statistic
of $\chi^2=537$ for 429 degrees of freedom (or $\chi^2_{red}$=1.25), compared
to the temperature of $1.22\pm0.19$~keV for a $\chi^2=415$ for 389 degrees of freedom (or $\chi^2_{red}$=1.08).
We therefore conclude that an accurate subtraction of point sources and unrelated
sources of diffuse emission is crucial to obtain an accurate measurement of the
temperature profile, especially in regions of low-surface brightness such as those
near the virial radius.

Changes in the instrument calibration affect the measurement of temperatures.
We therefore repeat the same data reduction and spectral analysis  using the latest 
software and calibration database available at time of writing (CIAO 4.4 and CALDB 4.5.1) for
the longest observation (ID 6880),
and obtain a new temperature profile for the same regions as reported in Table~\ref{tab:kT-err}.
In the outermost two regions, we measure a temperature of 3.04$\pm0.69$~keV (330-450") and 
1.23$\pm$0.21~keV (450-600"), well within the 1-$\sigma$ confidence intervals of the measurements
using the older calibration (3.40$\pm$0.76 and 1.22$\pm$0.19 respectively, also in agreement with the
values of Table ~\ref{tab:kT-err} obtained from the combination of all exposures). 
The temperature of the inner regions are also always within 1-$\sigma$
of the results obtained with the earlier calibration, and we therefore conclude that changes in the instrument
calibration do not affect significantly our results.

\section{Measurement of masses and gas mass fraction}

We fit the surface brightness and the temperature profiles with
the \cite{vikhlinin2006} model. The electron density is modelled
with a double-$\beta$ profile modified by a cuspy core component and an exponential cutoff at large radii,
for a total of eleven model parameters;
 the temperature has both a cool-core component to follow
the cooler gas in the core, and a decreasing profile at large radii, for an additional
nine parameters. For our analysis, we follow  \cite{vikhlinin2006} and fix the
$\gamma$=3.0 parameter, and do not use the cuspy-core component ($\alpha=0$)
or the second $\beta$-model component, so that the density is modelled by just 
one $\beta$-model with an exponential cutoff, for just four free parameters
(core radius $r_c$, exponent $\beta$, scale radius $r_s$ and exponential cutoff
exponent $\epsilon$, see Table~\ref{tab:vikh-fit}). 
For the temperature profile, we fix the parameter $a=0$, and the
remaining eight parameters are reported in Table~\ref{tab:vikh-fit}.

We use a Monte Carlo Markov chain (MCMC) method that we used in
previous papers \citep[e.g.,][]{bonamente2004,bonamente2006}.
The MCMC analysis consists of a projection of the three-dimensional
models and a comparison 
of the projected surface brightness and temperature profiles,
and results in simultaneous estimation
of the posterior distributions of all model paramters. Uncertainties in the  parameters are obtained
from the posterior distributions, with 1-$\sigma$ errors  assigned using the 68.3\%
confidence interval around the median of the distribution. 

The gas mass is directly calculated from the electron density
model parameters via
\begin{equation}
M_{gas}(r) = m_p \mu_e \int_0^r n_e(r) 4 \pi r^2 dr
\end{equation}
and the total gravitational mass via the equation of
hydrostatic equilibrium,
\begin{equation}
M(r) = - \frac{kT(r) r}{\mu_e m_p G} \left(\frac{d \ln n_e}{d \ln r} + \frac{d \ln kT}{d \ln r} \right),
\label{eq:hse}
\end{equation}
where $m_p$ is the proton mass, $\mu_e \simeq 1.17$ the mean electron molecular weight, and 
$G$ the gravitational constant. The total density of matter is simply obtained via
\begin{equation*}
\rho(r) = \frac{1}{4 \pi r^2} \frac{d M(r)}{dr}
\end{equation*}
and therefore can be obtained via a derivative of the mass profile.
In Equation~\ref{eq:hse}, the term $A = d \ln n_e/d \ln r + d \ln kT/d \ln r$
and its first derivative are always negative, as is $d kT(r)/d r$ at large radii.
Therefore, the density can be rewritten as
\begin{equation}
\rho(r) = - \frac{1}{4 \pi r^2 \mu_e m_p G} \left[ kT \left( A +r \frac{dA}{dr} \right)  + r A \frac{d kT}{dr} \right] 
\label{eq:density}
\end{equation}
in which the only negative term  is the one containing $A \cdot d kT(r)/d r$, while
the other two terms remain positive out to large radii.

\subsection{Modelling of the Chandra data out to the virial radius}
\label{sec:hse}
\begin{figure*}
\centering
\includegraphics[width=2.25in, angle=-90]{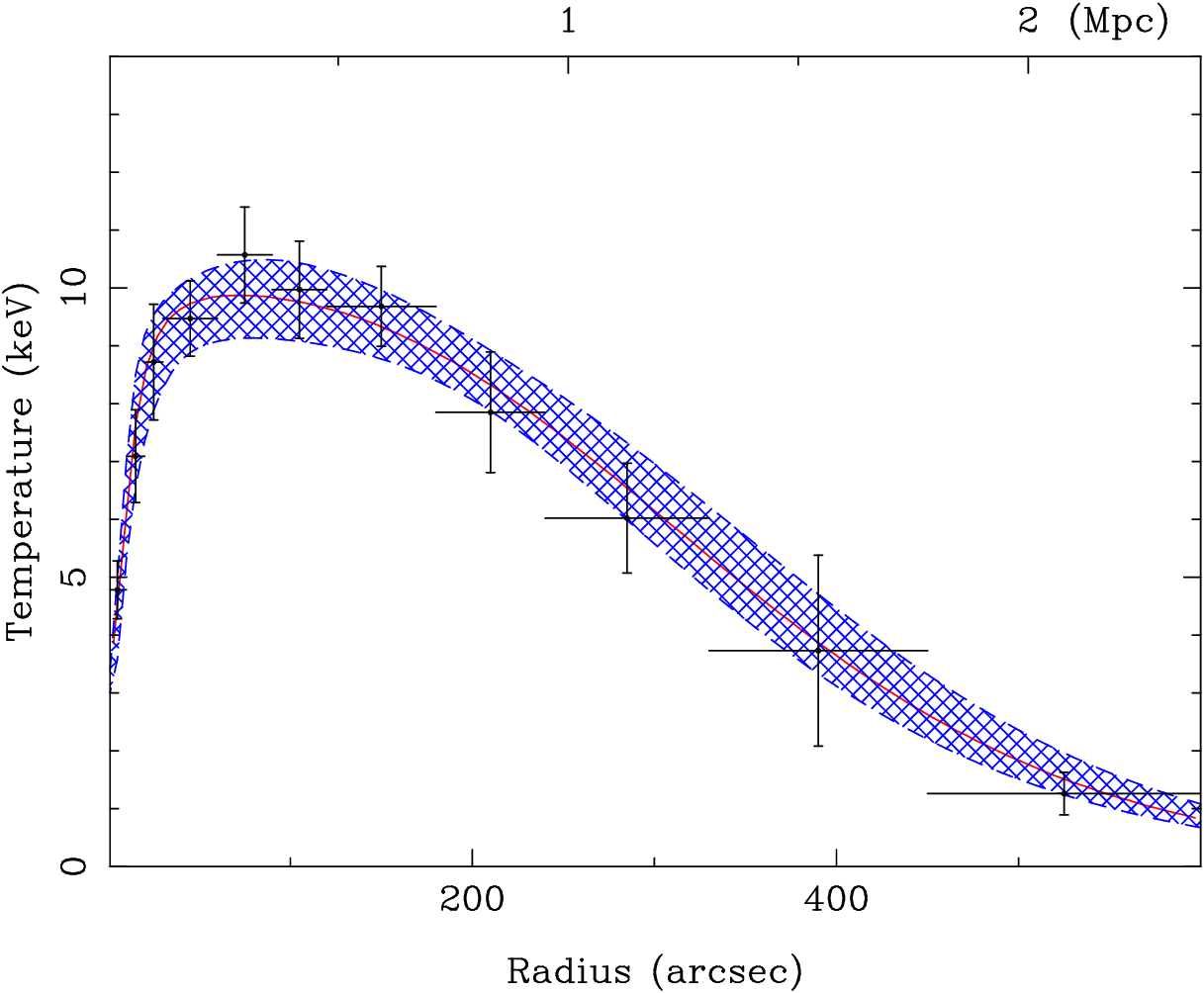}
\includegraphics[width=2.25in, angle=-90]{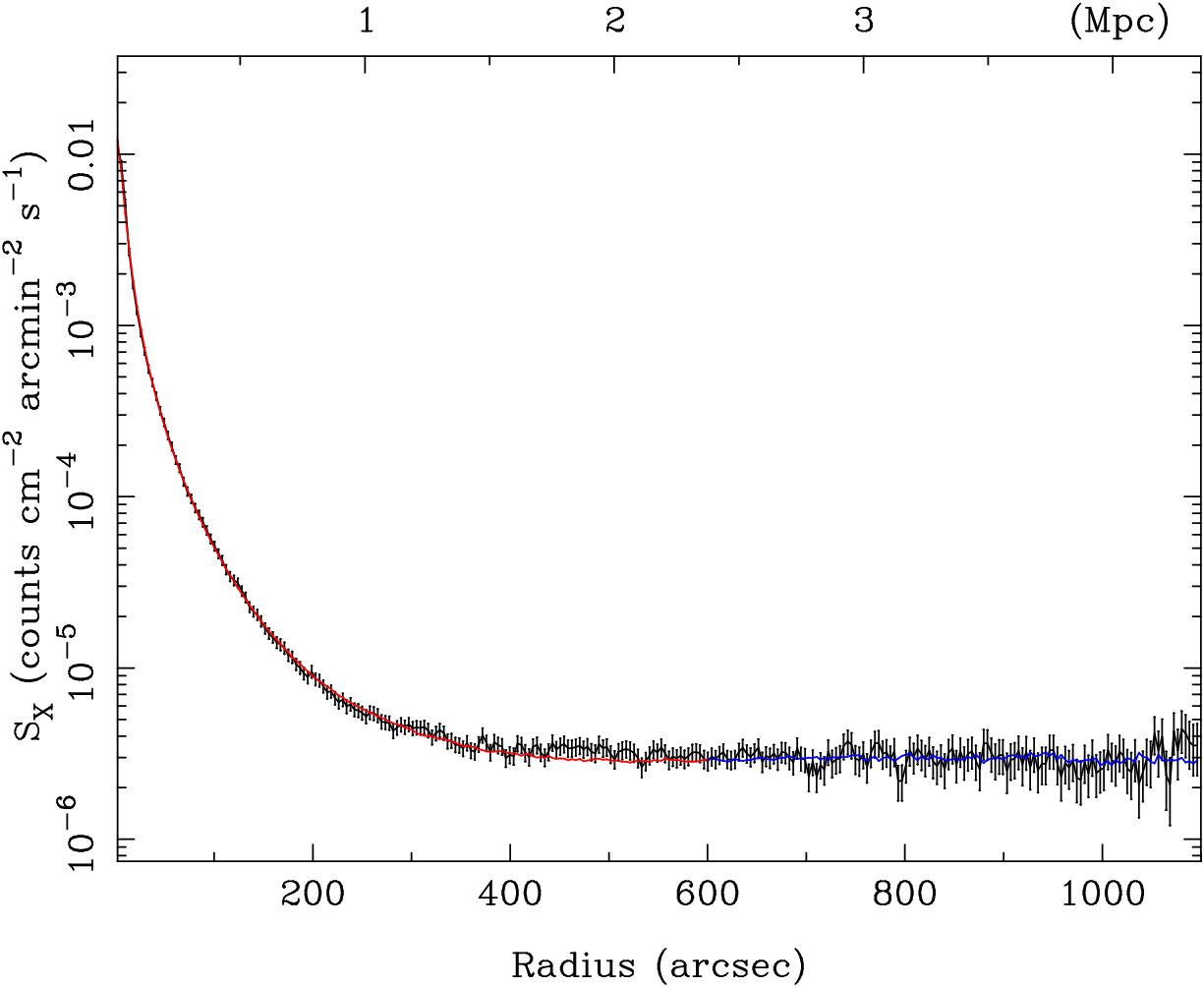}
\caption{Left: Best-fit Vikhlinin model for the projected temperature
profile out to 600", with 90\% confidence intervals. Right:  Best-fit
Vikhlinin model to the 0.7-2 keV surface brightness (model+background) profile.
Emission beyond 600" is statistically
consistent with the background, in  blue is the extrapolation out to 1100". 
Prior removal of the stowed background
caused the lower backgroud  level in Figure~\ref{fig:Sx}.
}
\label{fig:kt-0-600-fit}
\end{figure*}

\begin{figure*}
\centering
\includegraphics[width=2.3in, angle=-90]{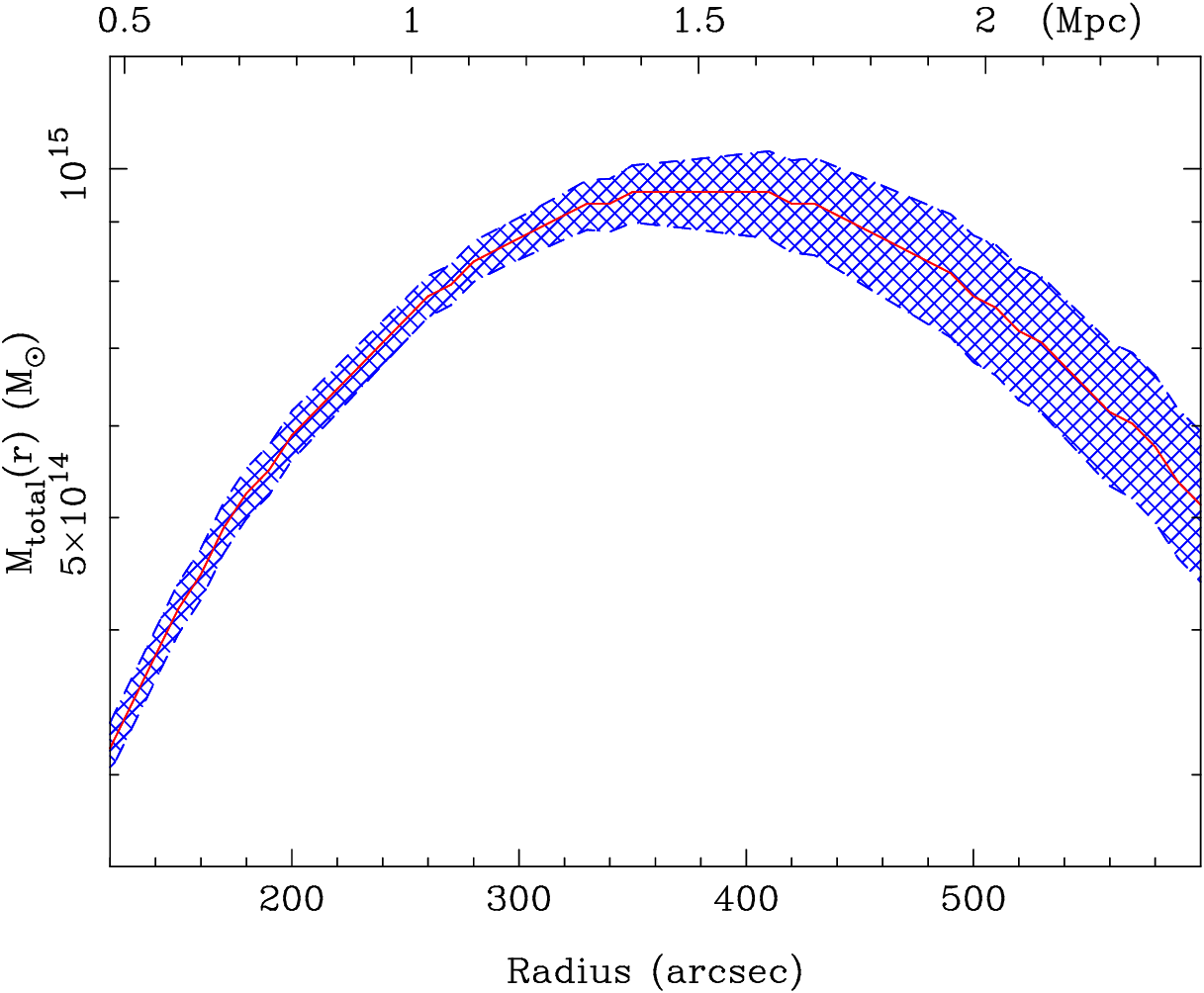}
\includegraphics[width=2.3in, angle=-90]{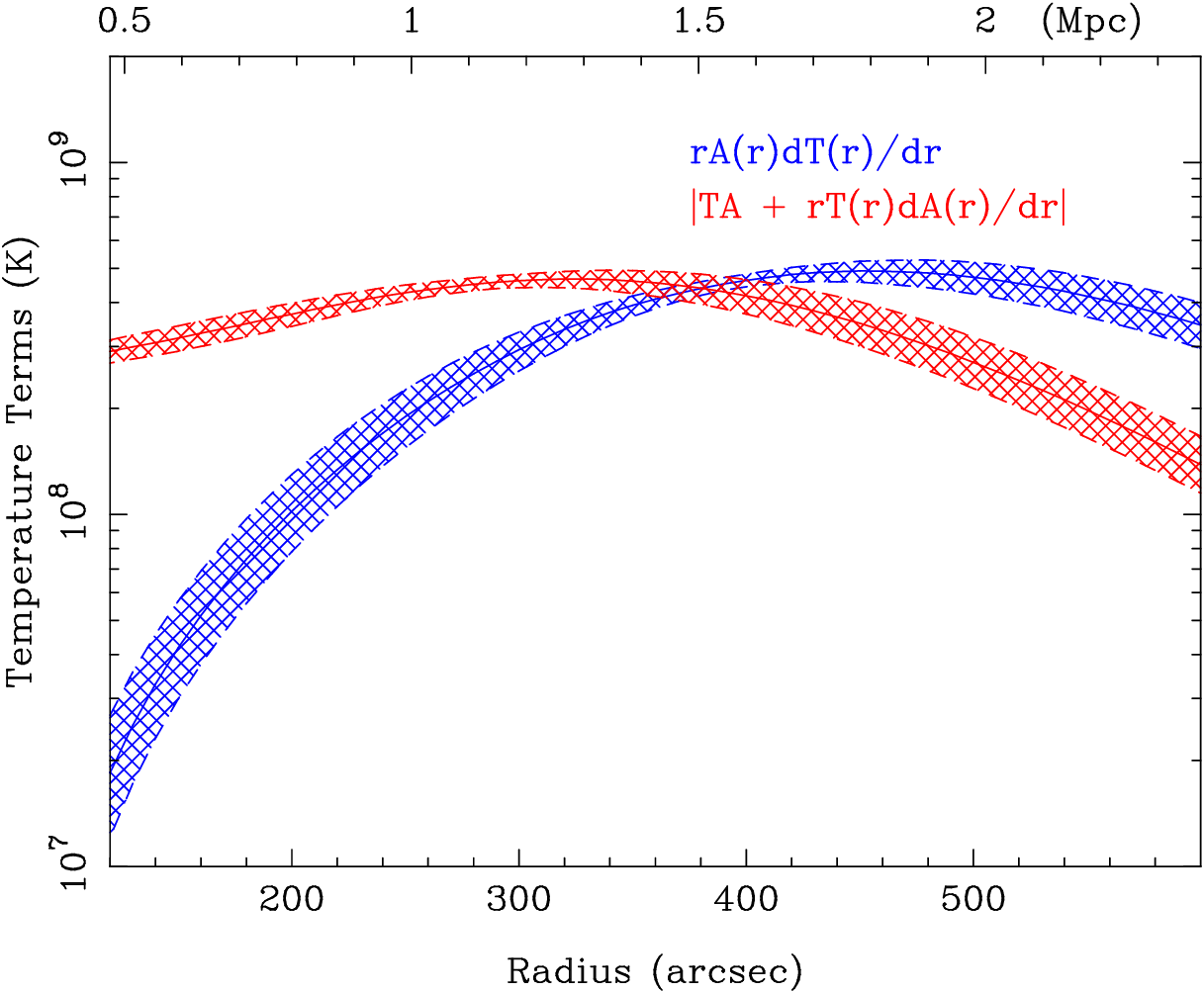}
\caption{Mass profile using data out to 600" and the temperature fit of Figure~\ref{fig:kt-0-600-fit},
and the radial distribution of the positive and negative terms in the density
equation (Equation~\ref{eq:density}).}
\label{fig:mass-0-600}
\end{figure*}

The \cite{vikhlinin2006} model provides a satisfactory fit
out to the outermost radius of 600";
Figure~\ref{fig:kt-0-600-fit} shows the best-fit models to the temperature 
and surface brightness profiles,
best-fit parameters
of the model are reported in Table~\ref{tab:vikh-fit}.
The temperature profile measured by \chandra\ in Figure~\ref{fig:kt-0-600-fit}
is so steep that it causes the total matter density $\rho(r)$ to become \emph{negative}
at approximately 400", indicating that the temperature profile 
cannot originate from gas in hydrostatic equilibrium. 
The situation is illustrated in Figure~\ref{fig:mass-0-600}, where the relevant terms
of Equation~\ref{eq:density} are plotted individually; 
 the density inferred from hydrostatic equilibrium becomes negative 
where the
negative term crosses the positive ones,
 and the mass profile has a negative slope beyond that point. These fit parameters
therefore lead to an unacceptable situation, and responsibility for this inconsistency
can be attributed to an overly steep temperature profile, with a drop by a factor of
ten between approximately 1.5' to 10'.

\begin{table*}
\centering
\caption{Best-fit parameters for the Vikhlinin model using \chandra\ data out to 330"}
\label{tab:vikh-fit}
\begin{tabular}{cccccccccc}
\hline
$n_{e0}$ & $r_{c}$ & $\beta$ & $r_{s}$ & $\epsilon$ & $n_{e02}$ & $\gamma$ & $\alpha$ & $\chi^{2}_{tot} \textrm{(d.o.f.)}$\\
(10$^{-2} $cm$^{-3}$) & (arcsec) & & (arcsec) & & & & & &\\
\hline
\multicolumn{10}{c}{Using \chandra\ data out to 330"}\\
$9.602\pm^{0.488}_{0.415}$ & $6.743\pm^{0.373}_{0.403}$ & $0.498\pm^{0.009}_{0.009}$ & $119.8\pm^{13.3}_{13.4}$ & $1.226\pm^{0.098}_{0.097}$ 
& 0.0 & 3.0 & 0.0 & \nodata\\ 
\hline
\multicolumn{10}{c}{Using \chandra\ data out to 600"}\\
$9.763\pm^{0.447}_{0.450}$ & $6.346\pm^{0.385}_{0.343}$ & $0.488\pm^{0.009}_{0.009}$ & $96.44\pm^{9.55}_{8.67}$ & $1.067\pm^{0.075}_{0.079}$
& 0.0 & 3.0 & 0.0 & \nodata\\
\hline
\hline
$T_{0}$ & $T_{min}$ & $r_{cool}$ & $a_{cool}$ & $r_{t}$ & $a_{t}$ & $b_{t}$ & $c_{t}$ &\\
(keV) & (keV) & (arcsec) & & (arcsec) & & & & &\\
\hline
\multicolumn{10}{c}{Using \chandra\ data out to 330"}\\
$38.25\pm^{19.63}_{17.23}$ & 3.0 & $92.48\pm^{52.63}_{40.52}$ & 1.0 & $257.5\pm^{143.0}_{66.72}$ & 0.0 & $1.024\pm^{0.426}_{0.283}$ & 2.0 &  39.0 (83)\\
\hline
\multicolumn{10}{c}{Using \chandra\ data out to 600"}\\
$10.17\pm^{0.85}_{0.60}$ & 3.0 & $11.82\pm^{3.61}_{2.29}$ & $1.924\pm^{0.802}_{0.568}$ & 600.0 & 0.0 & $2.800\pm^{0.224}_{0.210}$ & 10.0 & 106.4 (154)\\
\hline
\end{tabular}
\end{table*}


\begin{table*}
\centering
\caption{Masses Calculated using \chandra\ data out to 330", and Extrapolated out to $r_{100}$}
\label{tab:vikh-masses}
\begin{tabular}{ccccc}
\hline
$\Delta$ & $r_{\Delta}$ & $M_{gas}$ & $M_{total}$ & $f_{gas}$\\
         & (arcsec)     & $\times 10^{13}~M_{\odot}$ & $\times 10^{14}~M_{\odot}$ & \\
\hline
2500 & $164.9\pm^{4.1}_{3.9}$ & $4.70\pm^{0.15}_{0.14}$ & $5.03\pm^{0.38}_{0.35}$ & $0.093\pm^{0.004}_{0.004}$\\
500  & $326.6\pm^{7.1}_{6.9}$ & $10.75\pm^{0.23}_{0.23}$ & $7.80\pm^{0.52}_{0.49}$ & $0.138\pm^{0.006}_{0.006}$\\
200  & $453.3\pm^{15.2}_{15.1}$ & $15.36\pm^{0.48}_{0.48}$ & $8.35\pm^{0.86}_{0.81}$ & $0.184\pm^{0.014}_{0.012}$\\
100  & $570.9\pm^{26.6}_{25.3}$ & $19.53\pm^{0.84}_{0.82}$ & $8.34\pm^{1.22}_{1.06}$ & $0.234\pm^{0.024}_{0.022}$\\
\hline
\hline
\end{tabular}
\end{table*}


The results presented in this section provide
evidence  that the gas detected by \chandra\ near the virial radius is
\emph{not} in hydrostatic equilibrium, and a number of theoretical
studies do in fact suggest that  beyond \rfive\ the intergalactic
plasma is not supported solely by thermal pressure \citep[e.g.][]{lau2009}.
\suzaku\ has reported the measurement of emission near the
virial radius for several clusters, including \emph{Abell~1413},
\emph{Hydra~A}, \emph{Perseus}, \emph{PKS0745-191}, \emph{Abell~1795}, \emph{Abell~1689}
and \emph{Abell~2029}
\citep{hoshino2010,sato2012,simionescu2011,george2009, bautz2009,kawaharada2010,
walker2012a,walker2012b}. Some of these results do in fact report an apparent
decrease in total mass with radius \citep[e.g.]{george2009,kawaharada2010} 
and lack of hydrostatic equilibrium at large radii \citep[e.g.][]{bautz2009},  similar
to the results presented in this paper.
Temperature profiles measured by \suzaku\ typically do not feature as extreme a temperature drop
as the one reported in Figure~\ref{fig:kt-0-600-fit}, i.e., a factor of nearly 10
from peak to outer radius, although in some cases the drop of temperature from the peak
value to that at \rtwo\ is consistent with the one reported in this paper.

\subsection{Modelling of the Chandra data out to \rfive}
\label{sec:r500}
The steepening of the radial profile beyond 400" is driven by the
temperature of the last datapoint beyond $r_{200}$. 
%
We also model the surface brightness and temperature profiles of the
\chandra\ data out to only 330", or approximately \rfive, and find the
best-fit \cite{vikhlinin2006} model for the temperature profile
reported  in Figure~\ref{fig:kT-0-330} and Table~\ref{tab:vikh-fit}.
We measure a gas mass fraction of $f_{gas}(r_{500})=0.138\pm0.006$;
if we add the mean stellar fraction as measured by either \cite{giodini2009}
($f_{\star}=0.019\pm0.002$) or by \cite{gonzales2007} ($f_{\star}\simeq0.012$)
assuming $M(r_{500})=7.1\times 10^{14}$ $M_{\odot}$,
we find that \a1835\ has an average baryon content within  \rfive\
that is consistent with the cosmic abundance of $\Omega_b/\Omega_M=0.167\pm0.007$ \citep{komatsu2011}
at the 2-$\sigma$ level. As is the case in most clusters, especially relaxed
ones, the radial distribution of the gas mass fraction increases with radius
\citep[e.g.,][]{vikhlinin2006}.

We use this modelling of the data to measure \rfive, and to provide estimates
for \rtwo\ and the virial radius.
The extrapolation of this model to 600" now falls above the measured temperature profile,
and the mass profile using hydrostatic equilibrium is monotonic.
This best-fit model is marginally compatible with the assumption of hydrostatic equilibrium.
In fact, Table~\ref{tab:vikh-masses} shows that the extrapolated  mass profile
flattens around \rtwo, with virtually no additional mass being necessary
beyond this radius to sustain the hot gas in hydrostatic equilibrium.
Moreover, between \rfive\ and \rtwo, all of the gravitational mass is accounted
by the hot gas mass, i.e., \emph{no} dark matter is required beyond \rfive.
This extrapolation of the $\leq$~\rfive\ data to the virial radius therefore
leads to a dark matter halo that is much more concentrated than
the hot gas.

\section{Entropy profile and convective instability at
large radii}
\label{sec:entropy}
The Schwarzschild
criterion for the onset of convective instability is given by the
condition of buoyancy of an infinitesimal blob of gas that is displaced by an
amount $dr$, $d \rho_{blob} < d \rho$,
where $\rho_{blob}$ is the density of the displaced blob, assumed to attain pressure
equilibrium with the surrounding, and $\rho$ is the density of ambient medium.
If the blob is displaced adiabatically, using pressure $P$ and entropy $s$ as the
independent thermodynamic variables in the derivatives of $\rho_{blob}$ and $\rho$, 
the buoyancy condition gives 
\begin{equation}
\left. \frac{\partial \rho}{\partial s} \right|_{p} ds > 0
\label{eq:buoyancy}
\end{equation}
as condition for convective instability, i.e., a blob that is displaced radially outward will
find itself in a medium of higher density and continue to rise to larger radii. Since
$({\partial \rho}/{\partial s})_P=-\rho^2 (\partial T / \partial P)_s<0$ (material
is heated upon adiabatic compression),  Equation~\ref{eq:buoyancy} simply
reads that \emph{a radially decreasing entropy profile is convective unstable}.

\begin{figure*}
\centering
\includegraphics[width=2.3in, angle=-90]{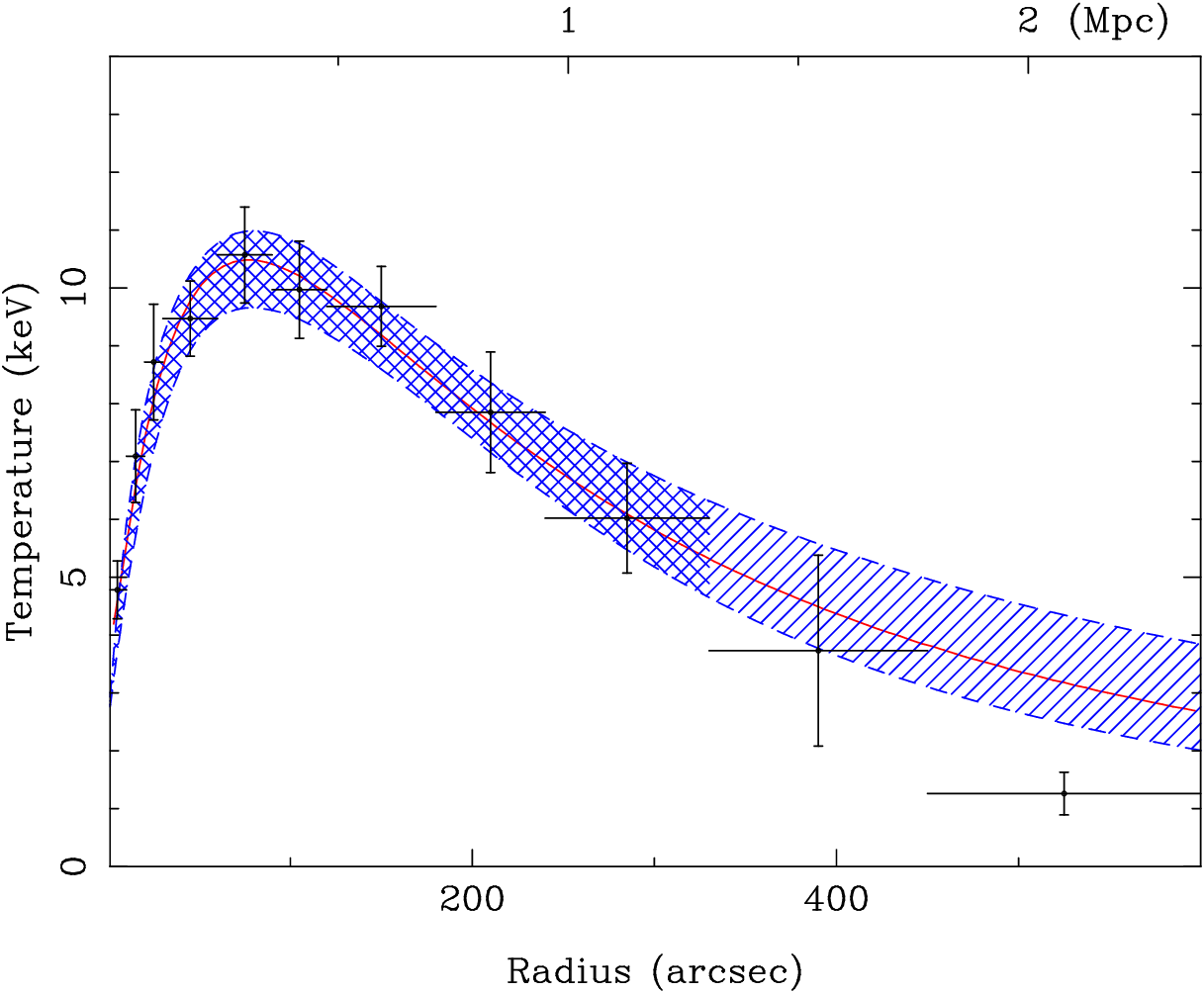}
\includegraphics[width=2.3in, angle=-90]{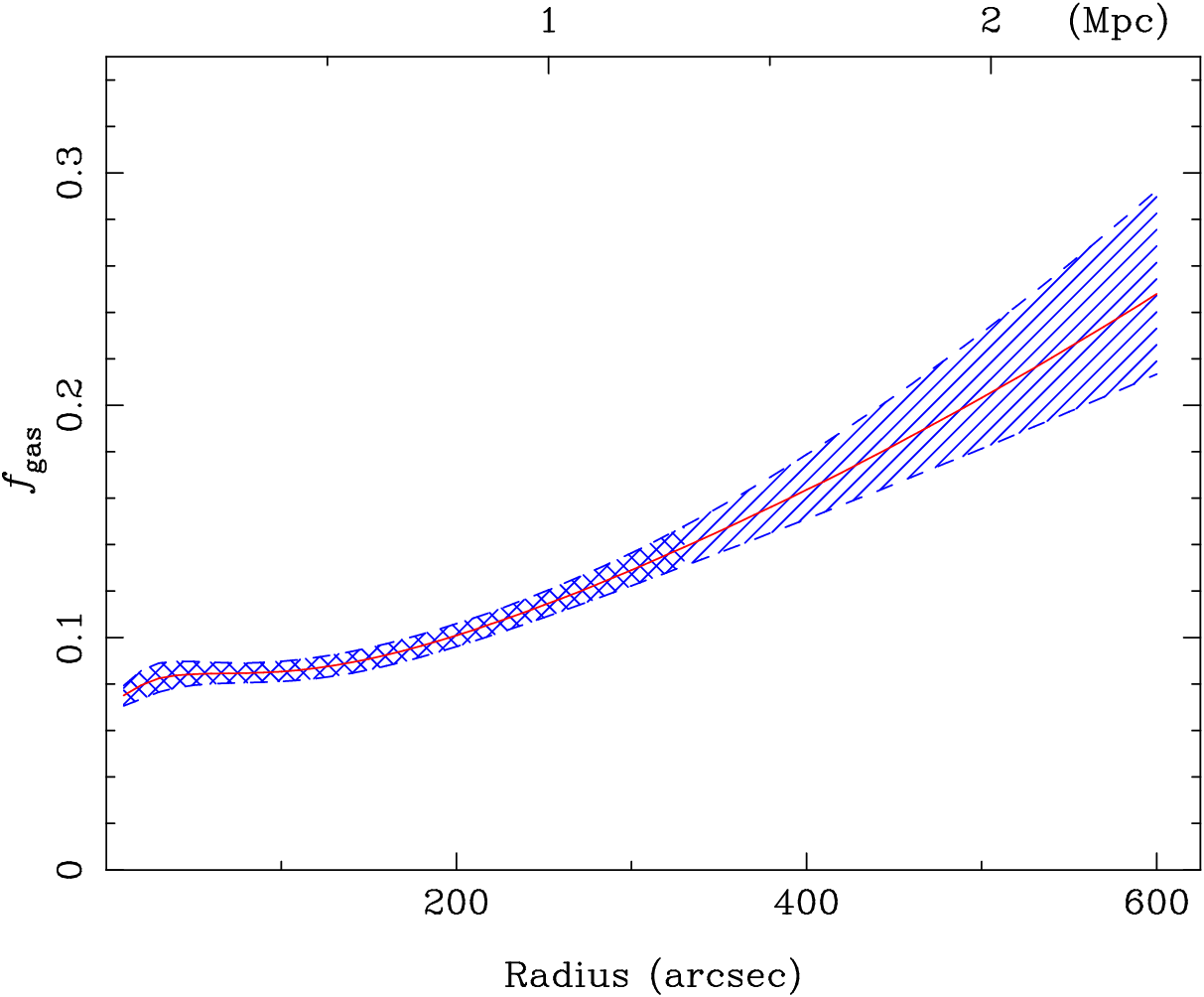}
\caption{
Temperature and gas mass fraction profiles measured from a fit to the \chandra\ data out to 330", and extrapolation of the
best-fit model out to 600".}
\label{fig:kT-0-330}
\end{figure*}

\begin{figure*}
\centering
\includegraphics[width=2.3in, angle=-90]{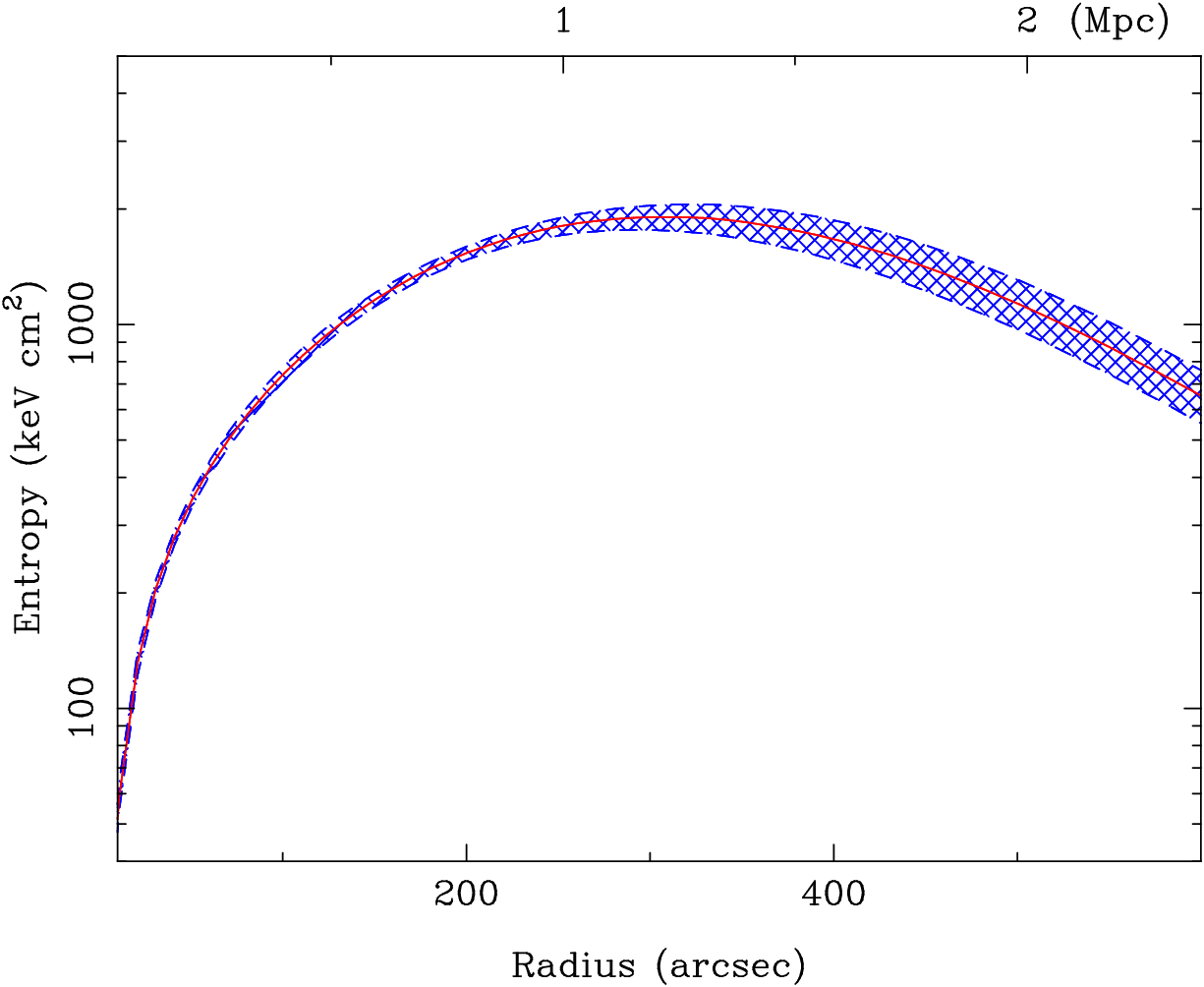}
\includegraphics[width=2.3in, angle=-90]{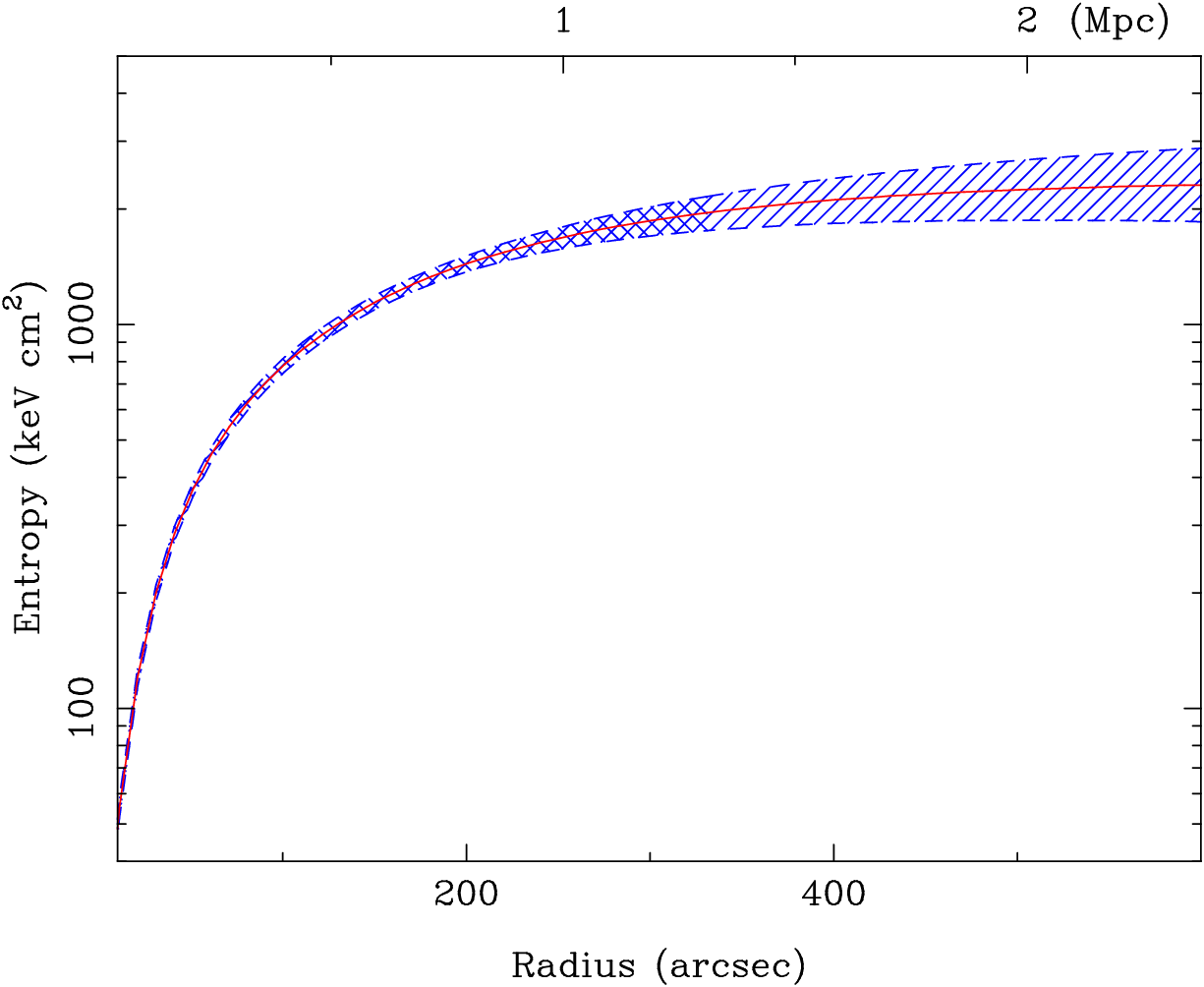}
\caption{
Deprojected entropy profiles using the full \chandra\ data out to 600" (left, see Section~\ref{sec:hse}),
and using only data out to \rfive\ (right, see Section~\ref{sec:r500}).}
\label{fig:entropy-profile}
\end{figure*}

An ideal gas has an entropy of
\begin{equation}
S = \nu R \left( \frac{3}{2} \ln T - \ln \rho + C\right)
\end{equation}
where $\nu$ is the number of moles, $R$ is the gas constant, and $C$ is a constant.
In astrophysical applications, it is customary \citep[e.g.][]{cavagnolo2009} to use a definition
of entropy that is related to the thermodynamic entropy by an operation of
exponential and a constant offset,
\begin{equation}
S = \frac{kT}{n_e^{2/3}},
\label{eq:entropy}
\end{equation}
The entropy $S$ defined by Equation~\ref{eq:entropy} has
units
of keV cm$^{2}$, and it is required to be radially increasing to maintain convective equilibrium.
Numerical simulations
indicate that entropy outside the core is predicted to increase with radius approximately 
as $r^{1.1}$ or $r^{1.2}$  \citep{voit2005,tozzi2001}.
In Figure~\ref{fig:entropy-profile} we show the radial profile of the entropy out to
the outer radius of 10 arcmin, with a significant decrease at large radii that indicates
an incompatibility of the best-fit model with convective equilibrium. For comparison,
we also show the entropy profile measured using the modelling of the data
out to only \rfive, as described in Sec.~\ref{sec:r500}. This entropy profile
uses the shallower temperature profile of Figure~\ref{fig:kT-0-330}, and its
extrapolation to larger radii remains non-decreasing, i.e., marginally consistent
with convective equilibrium.

The Schwarzschild criterion 
does not apply in the presence of a magnetic field. For typical
values of the thermodynamic quantities of the ICM, the electron and ion gyroradii are
several orders of magnitude smaller than the mean free path for Coulomb collisions
\citep[e.g.][]{sarazin1988}, even for a magnetic field of order 1 $\mu G$, and therefore
diffusion takes place primarily along field lines \citep[e.g.][]{chandran2007}.
There is strong evidence of magnetic
fields in the central regions of clusters \citep[e.g., radio halos, ][]{venturi2008,cassano2006},
though it is not clear whether magnetic fields are ubiquitous
near the virial radius, as in the case of Abell~3376 \citep{bagchi2006}.
In the presence of magnetic fields, \cite{chandran2007} has shown that the
condition for convective instability is simply $dT/dR<0$.

  
The \chandra\ data out to the virial radius therefore indicate
that the ICM is convectively unstable, regardless of the
presence of a magnetic field. In fact, in the absence of magnetic
fields near the virial radius, Figure~\ref{fig:entropy-profile} shows that \a1835\
fails the standard Schwarzschild criterion, i.e., the entropy decreases with radius;
in the presence of magnetic fields, the negative gradient in the temperature profile alone
is sufficient for the onset of convective instability 
\citep[e.g., as discussed by ][]{chandran2007}.
  Convective instabilities would carry hotter
gas from the inner regions towards the outer region within a few sound crossing
times. As shown by \cite{sarazin1988}, the sound crossing time for a 10~keV
gas is $\sim 0.7$~Gyr for a 1~Mpc distance,
and an unstable temperature gradient such as that of Figure~\ref{fig:kt-0-600-fit}
would be flattened by convection within a few Gyrs.
Convection could in principle also result in an additional pressure gradient
due to the flow of hot plasma to large radii, which can in turn help support the gas
against gravitational forces. 

\section{Discussion and interpretation}
In this paper we have reported the detection of X-ray emission in \a1835\ with \chandra\ that extends out to approximately
the cluster's virial radius. The emission can be explained by the presence of a cooler
phase of the plasma that is dominant at large radii, possibly linked to the infall
of gas from large-scale filamentary structures. We also investigate the effects of clumping of the gas 
at large radii, and conclude that in principle a radial gradient in the clumping factor of the hot ICM
can explain the apparent flattening of the entropy profile and the turn-over of the mass profile.
\subsection{Detection of X-ray emission out to the virial radius}
The detection of X-ray emission out to a radial distance of 10 arcmin, or approximately 2.4~Mpc,
indicates the presence of diffuse gas out to the cluster's virial radius.
This is the first detection of gas out to the virial radius with \chandra, matching
other detections obtained with \suzaku\ for nearby clusters 
\citep[e.g.][]{akamatsu2011,walker2012a,walker2012b,simionescu2011,burns2010,kawaharada2010,
bautz2009,george2009}.
Despite its higher background, \chandra\ provides a superior angular resolution to image and remove emission from unrelated sources.
As can be seen from Figure~\ref{fig:a1835}, there are approximately 100 point-like sources that were automatically
detected and removed, and we were also able to identify two low-mass clusters that are likely associated with \a1835.
\chandra\ therefore has the ability to constrain the emission of clusters to the virial radius, especially for higher-redshift
cool-core clusters for which the \suzaku\ point-spread function would cause significant contamination from the 
central signal to large radii.

It is not easy to interpret the emission at the outskirts as an extension 
of the hot gas detected at radii $\leq$~\rfive. In fact, as shown in Section~\ref{sec:hse}, the steepening of the
temperature profile is incompatible with the assumption of
hydrostatic equilibrium at large radii. 
We also showed in Section~\ref{sec:entropy} that
the gas has a negative entropy gradient beyond this radius, rendering it convectively unstable. 
Therefore, if the temperature profile of Figure~\ref{fig:kt-0-600-fit} originates from
a single phase of the ICM, convection would transport hotter gas towards the outskirts, flattening
the temperature profile within a few Gyrs. Cooling of the gas by thermal radiation cannot be
responsible for off-setting the heating by convection, since the cooling time
 ($t_{cool} \sim kT^{1/2} n_e^{-1}$) is longer at the outskirts than in the peak-temperature regions
due to the higher density.

\subsection{Warm-hot gas towards the cluster outskirts}
A possible interpretation for the detection of emission near the virial radius and its
steep temperature profile is the presence of a separate phase at the cluster outskirts
that is not in hydrostatic equilibrum with the cluster's potential.
In this case, cooler gas may be the result of infall from filamentary structures that
feed gas into the cluster, and the temperature of this \emph{warm-hot} gas may in fact be
lower than that shown in Figure~\ref{fig:kt-0-600-fit} 
(i.e., $kT \sim 1.25$~keV for the region $\geq 450$") if
this gas lies in projection against the tail end of the hotter ICM.

We estimate the mass of this putative warm-hot gas assuming that all of the
emission from the outermost region is from a uniform density gas
seen in projection. This assumption may result in an overestimate
of the emission measure; in fact,  the extrapolation of the gas density profile 
in the hydrostatic or convective scenarios may yield a significant amount of
emission in the last radial bin. 
We were unable to perform a self-consistent modelling
of the emission  in the full radial range, since the low signal-to-noise
ratio does not allow a two-phase modelling in the last radial bin.
In this simple uniform density warm-hot gas scenario, 
the gas is in a filamentary structure
of length $L$ and area $A=\pi(R_{out}^2-R_{in}^2)$, where
$R_{out}=600$" and $R_{in}=450"$; this is the same model
also considered in \cite{bonamente2005} for the cluster \emph{Abell~S1101}.
Since the length $L$ of the filament along the sightline is unknown,
we must either assume $L$ or the electron density $n_e$, and 
estimate the mass implied by the detected emission.
The emission integral for this region is proportional to
\begin{equation}
K = \frac{10^{-14}}{4 \pi D_A^2 (1+z)^2} n_e^2 V,
\end{equation}
where $K$ is measured in XSPEC from a fit to the spectrum, $D_A$ is
the angular distance in cm, $z$ is the cluster redshift, and 
the volume is $V=A \times L$. For this estimate we assume 
for simplicity that
the mean atomic weights of hydrogen and of the electrons are
the same, $\mu_e=\mu_H$.
Using the best-fit spectral model with $kT=1.26\pm0.16$ keV,
we measure $K=1.05\pm 0.13 \times 10^{-4}$. If we assume a filament of
length $L=10$~Mpc, then the average density is $n_e=2.4\pm0.3$~cm$^{-3}$,
and the filament mass is $4.6\pm0.6 \times 10^{13}$~$M_{\odot}$.
Alternatively, a more diffuse filament gas of $n_e=10^{-5}$~cm$^{-3}$
would require a filament of length $L=58\pm8$~Mpc, with
a mass of $1.1\pm0.2\times 10^{14}$~$M_{\odot}$, comparable to the
entire hot gas mass within \rtwo. The fact that a lower density gas
yields a higher mass is given by the fact that, for a measured value of $K$
we obtain $n_e \propto L^{-1/2}$, and therefore the mass is proportional to $L^{1/2}$.
For comparison, the gas mass for this shell inferred from the standard
analysis, i.e.,  assuming that the gas is in the shell itself,
is $\sim 3\times 10^{13}$~$M_{\odot}$, as can be also seen from Table~\ref{tab:vikh-masses}.

If the gas is cooler, then the mass budget would increase further.
In fact, the bulk of the emission from cooler gas falls outside of the \chandra\ bandpass,
and for a fixed number of detected counts the required emission integral increases.
We illustrate this situation by fitting the annulus to an emission
model with a fixed value of $KT=0.5$~keV, which result in a value
of $K=1.88\pm 0.24 \times 10^{-4}$ (the fit is significantly poorer, with
$\Delta \chi^2=+10$ for one fewer degree of freedom). 
Accordingly, the filament mass estimates would be increased 
approximately by a factor of two. 

A warm-hot phase at $T\leq 10^7$~K is expected to be a significant reservoir of baryons
in the universe \citep[e.g.][]{cen1999,dave2001}. Using the \rosat\ soft X-ray 
Position Sensitive Proportional Counter (PSPC)  detector --  better suited to
detect the emission from sub-keV plasma --  we
have already reported 
the detection of a large-scale halo of emission around the \emph{Coma} cluster out to $\sim$~5 Mpc, well beyond the
cluster's virial radius \citep{bonamente2003,bonamente2009}. 
It is possible to speculate that the high mass of \a1835, one
of the most luminous and massive clusters on the \emph{Bright Cluster Survey} sample \citep{ebeling1998},
is responsible for the heating of the infalling gas to temperatures that makes it
detectable by \chandra, and that other massive clusters may therefore provide
evidence of emission to the virial radius with the \chandra\ ACIS detectors.  
The infall scenario is supported by the \emph{Herschel} observations
of \cite{pereira2010}, who measure a galaxy velocity distribution for \a1835\
that does not appear to decline at large radii as in most of the other clusters
in their sample. A possible interpretation for their data is the presence of a 
surrounding filamentary structure  that is infalling into the cluster.

\subsection{Effects of gas clumping at large radii}
Masses and entropy measured in this paper assume that the gas has a uniform density
at each radius.  To quantify the effect of departures from uniform density, we
   define the clumping factor $C$ 
as the ratio of density averages over a large region,
\begin{align}
C & = \frac{\langle n_e^2 \rangle}{\langle n_e \rangle^2}
\end{align}
with $C \geqslant 1$.
Clumped gas emits more efficiently than gas of uniform
density, 
and the same surface brightness $I$ results in a lower estimate for the gas density and mass,
\begin{equation}
I \propto \int <n_e^2> dl = \int <n_e>^2 C dl,
\end{equation}
where $l$ is a distance along the sightline.
From Figure~\ref{fig:entropy-profile} we see that the entropy drop from
approximately 400" to 600" would be offset by a decrease in $n_e^{2/3}$ by a factor
of 3, or a decrease in $n_e$ by a factor of 5. We therefore suggest that a clumping
factor of $C \simeq 25$ at 600" would in principle be able to provide
a flat entropy profile, and even higher clumping factors would provide
an increasing entropy profile in better agreement with theory \citep[e.g.][]{voit2005,tozzi2001}.
Numerical simulations by \cite{nagai2011} suggest  values of the clumping factor
$C \leq 3$ near \rtwo, with significantly higher clumping possible at larger radii. 
Use of the \cite{nagai2011}  model in the analysis of a large sample of galaxy clusters by \cite{eckert2012}
results in better agreement of observations with numerical simulations. 

Clumping can also affect the measurement of hydrostatic masses.
In particular, gas with an increasing radial profile of the clumping factor
could result in a steeper gradient of the density profile, when compared with what is measured assuming
a uniform density. According to Equation~\ref{eq:hse}, this 
would result in larger estimates of the hydrostatic mass, in principle able to reduce or entirely
offset the apparent decreas of $M(r)$ reported in Figure~\ref{fig:mass-0-600}.
We therefore conclude that a radial increase in the clumping of the gas can in principle
account for the apparent decrease of the mass profile and of the entropy profile
reported in this paper (Figures~\ref{fig:mass-0-600} and \ref{fig:entropy-profile}), and therefore
it is a viable scenario to interpret our \chandra\ observations.
Clumping of the gas at large radii has also been suggested based on \suzaku\ observations
\citep[e.g.,][]{simionescu2011}.

\section{Conclusions}
In this paper we have reported the detection of emission from \a1835\ with \chandra\
out to the cluster's virial radius. The cluster's surface brightness
is significantly above the background level out to a radius of
approximately 10 arcminutes, which correspond to $\sim$2.4 Mpc at the
cluster's redshift. We have investigated several sources of systematic
errors in the background subtraction process, and determined that the
significance of the detection in the outer region (450-600") is
$\geq 4.7$~$\sigma$, and the emission cannot be explained
by fluctuations in the background. Detection out to the virial
radius is also implied by the \xmm\ temperature profile
reported by \cite{snowden2008}.

The \chandra\ superior angular resolution made it straightforward to
identify and subtract sources of X-ray emission that are unrelated to the cluster. 
In addition to a large number of point sources, we have identified X-ray emission
from two low-mass clusters that were selected from the SDSS data,
MAXBCG J210.31728+02.75364 \citep{koester2007}
and WHL J140031.8+025443 \citep{wen2009}.
The two clusters have photometric and spectroscopic redshifts that make them
likely associated with \a1835. These are the only two
SDSS-selected clusters that are in the vicinity of \a1835.

The outer regions of the \a1835\ cluster have a sharp drop in the temperature
profile, a factor of about ten from the peak temperature. The sharp drop
in temperature implies that the hot gas cannot be in hydrostatic equilibrium, and
that the hot gas would be convectively unstable. A possible scenario to
explain the observations is the presence of \emph{warm-hot} gas
near the virial radius that is not in hydrostatic equilibrium with
the cluster's potential, and with a mass budget comparable to that
of the entire ICM. The data  are also consistent with an alternative scenario
in which a significant clumping of the gas at large radii is responsible
for the apparent negative gradients of the mass and entropy profiles
at large radii.

\bibliographystyle{mn2e}
\bibliographystyle{apj}



\label{lastpage}
\end{document}